\documentstyle[preprint,prb,aps]{revtex}
\begin{document}
\draft
\title{Magnetic anisotropies and general on--site Coulomb interactions
in the cuprates}
\author{O. Entin-Wohlman,$^1$ A. B. Harris,$^{1,2}$ and Amnon Aharony$^1$}
\address{1. School of Physics and Astronomy, Raymond and Beverly Sackler
Faculty of Exact Sciences,\\ Tel Aviv University, Tel Aviv 69978, Israel}
\address{2.  Department of Physics, University of Pennsylvania,
Philadelphia, Pennsylvania 19104-6369}
\date{\today}
\maketitle
\begin{abstract}
This paper derives the anisotropic superexchange interactions from
a Hubbard model for excitations within the copper 3d band and the oxygen
2p band of the undoped insulating cuprates. We extend the
recent calculation of Yildirim et al. [Phys. Rev. B {\bf VV}, pp, 1995]
in order to include the most general on--site Coulomb interactions
(including those which involve more than two orbitals)
when two holes occupy the same site.  Our general results apply when
the oxygen ions surrounding the copper ions form an octahedron
which has tetragonal symmetry (but may be rotated as in lanthanum
cuprate).  For the tetragonal cuprates we obtain an easy--plane
anisotropy in good agreement with experimental values.  We
predict the magnitude of the small in--plane anisotropy gap in
the spin--wave spectrum of YBa$_2$Cu$_3$O$_6$.
\end{abstract}

\section{Introduction}

The magnetic anisotropies of the family of compounds with structures
similar to that of La$_{2}$CuO$_4$ (LCO) has been a subject of much current
interest. \cite{ANIS1,ANIS3,ANIS2,ANIS4,ANIS5,PRL,PRB,ANIS6}
These materials are antiferromagnets, \cite{LCO} with a
dominant easy--plane anisotropy.\cite{PLANE}  The parent compound,
LCO, is orthorhombic in the regime of interest.  In this structure
the oxygen octahedra surrounding each copper rotate by a small angle
relative to their orientation in the tetragonal phase. This
phase was found to have an antisymmetric
Dzyaloshinskii--Moriya interaction, as allowed by the lack of
inversion symmetry about the center of a Cu-Cu bond.\cite{THIO}
More recently, members of this family which remain tetragonal,
such as Sr$_2$CuO$_2$Cl$_2$,\cite{VAKNIN} have been studied.
In many such compounds, for example, Sr$_2$CuO$_2$Cl$_2$,\cite{GREVEN}
Nd$_2$CuO$_4$,\cite{BOURGES} and Pr$_2$CuO$_4$\cite{SUMARLIN}
the gap in the spin--wave spectrum at zero wave vector due to
the easy--plane anisotropy has been found to be about 5meV,
just as in LCO.\cite{KEIMER}

Since these systems involve copper ions in a 3d$^9$ configuration
which have spin 1/2, the usually dominant mechanism of single--ion
anisotropy does not come into play.  Instead, anisotropy must be
due to the anisotropy of the superexchange interaction.
Microscopic derivations of the anisotropy energies have been
carried out \cite{ANIS1,ANIS3,ANIS2,ANIS4,ANIS5,PRL,PRB,ANIS6}
on the basis of Anderson's
theory \cite{ANDERSON} of kinetic superexchange, and
Moriya's extension \cite{MORIYA1} to incorporate spin--orbit
interactions. These have been mostly confined to the consideration
of the orthorhombic phase and were based on terms requiring the
existence of a distortion. As a result, these calculations produce
anisotropy energies which are proportional to the distortion angle
and which therefore vanish
in the tetragonal phase. However, since the experiments cited
above\cite{GREVEN,BOURGES,SUMARLIN,KEIMER} indicate that the magnitude
of the easy--plane anisotropy is independent of the structural
distortion, these theories, while giving correct information about
the Dzyaloshinskii-Moriya interaction, \cite{MORIYA1,MORIYA2}
did not provide a satisfactory
basis for understanding the easy--plane anisotropy.  Recognizing
this fact, Yildirim et al. \cite{PRL,PRB} undertook
an investigation of a model designed to calculate the anisotropy
of the superexchange interaction so as to account correctly for
the tetragonal symmetry of the lattice.  In their treatment all
five crystal field states of the copper 3d band (with a single hole
on each copper ion) were taken into account, as well as all the (occupied)
2p states of the oxygen ions.  As a result, the exchange interaction
associated with each Cu--Cu bond had biaxial exchange anisotropy.
(I. e. all three diagonal components of the exchange tensor were
in general different.)  In that work, a model of Coulomb interactions
was used which contained more terms than usual, but was still
not completely general.  Here we carry out the calculation
with completely general on--site Coulomb interactions necessary
to treat excited states with two holes on a single Cu ion,
or on a single oxygen ion, assuming tetragonal site symmetry.
As is known, \cite{GRIFF} such interactions can be parametrized
in terms of only three parameters, the Racah parameters, $A$,
$B$, and $C$, with $A \gg B$ and $A\gg C$.  In Refs. \onlinecite{PRL}
and \onlinecite{PRB} it was shown that for tetragonal symmetry
the exchange anisotropy vanishes if $B=C=0$.  Here we
obtain results which include all contributions to the
exchange anisotropy correct to first order in both $B$ and
$C$.  Compared to the previous work, we find an additional
contribution which slightly increases the biaxiality of the
anisotropy, without sensibly changing the easy--plane
anisotropy, which still agrees quite well with experimental
results.  Under certain approximations, our results can be
applied to the orthorhombic phase of LCO.  We should emphasize,
however, that our aim in this paper is to calculate only
the anisotropy of the exchange interaction.  This point
is discussed at the end of Sec. III.

Briefly, this paper is organized as follows.  In Sec. II we describe
the model for which we calculate the exchange anisotropy.  In
Sec. III we describe the perturbative calculations, Sec. IV
contains specific results for tetragonal symmetry, and in
Sec. V we discuss and briefly summarize our results.  We will attempt
to make this paper self--contained, but readers wishing more
details on the approach or the history of this problem are advised
to consult Ref. \onlinecite{PRB} (YHAE).  That reference also
contains the details of the spin--wave spectrum which results from
a calculation of the exchange anisotropy.

\section{The Model}

We start by considering the ground state of the CuO$_2$ plane, when
the kinetic energy is completely neglected.  In that case, there
is a single 3d hole on each copper ion and the oxygen 2p band is
completely full.  Since the spin of the Cu hole is arbitrary,
this ground state is $2^{N}$-fold degenerate.  The kinetic
superexchange interaction is obtained as the effective interaction
within this degenerate manifold when the kinetic energy is treated
perturbatively.  For this purpose we write the Hamiltonian as
\begin{equation}
\label{MICRO}
{\cal H}={\cal H}_{\rm Cu}+{\cal H}_{\rm O}+{\cal H}_{\rm hop}\ .
\end{equation}
Here, ${\cal H}_{\rm Cu}$ (${\cal H}_{\rm O}$) describes the
Hamiltonian of the copper (oxygen) ions and ${\cal H}_{\rm hop}$
is the kinetic energy (hopping between Cu and O ions).
To describe the Cu ions we work in a representation in which
the crystal field Hamiltonian is diagonal.
Diagonalization of the crystal field potential yields five spatial
$d$-states, denoted by $\mid\alpha \rangle$, with site energies
$\epsilon_{\alpha}$. Each such energy is doubly degenerate due
to spin $\sigma =\pm 1$.  In the systems under consideration,
the five $d$-states $\mid\alpha \rangle$ are determined by the tetragonal
symmetry. Even in the orthorhombic phase the site symmetry of the
copper ions can be taken to be tetragonal, since the crystal field
is primarily generated by the neighboring oxygen ions, which, to a
good approximation, still form an octahedron, albeit a rotated one.
For tetragonal symmetry we label these crystal field states as
$| \, 0 \rangle ~ \sim x^{2}-y^{2}$,
$| 1 \rangle ~ \sim 3z^{2}-r^{2}$, $|\, z \rangle ~ \sim xy$,
$|\, x \rangle ~ \sim yz$, and $|\, y \rangle ~ \sim zx$,
where the $z$-axis is perpendicular to the CuO$_{2}$ plane and
$|\, 0 \rangle$ is the lowest energy single--particle state.  Thus
\begin{eqnarray}
\label{CUHAM}
{\cal H}_{\rm Cu} &=& \sum_{i\alpha\sigma}\epsilon_{\alpha}
d_{i\alpha\sigma}^{\dagger} d_{i\alpha\sigma}
+ \frac{\lambda}{2}\sum_{i\alpha\beta}\sum_{\sigma\sigma '}
{\bf L}_{\alpha\beta}\cdot [\vec{\sigma}]_{\sigma\sigma '}
d_{i\alpha\sigma}^{\dagger} d_{i\beta\sigma '} \nonumber \\
&+& \frac{1}{2}\sum_{\alpha\beta\gamma\delta}
\sum_{i\sigma\sigma'}U_{\alpha\beta\gamma\delta}
d_{i\alpha\sigma}^{\dagger}d_{i\beta\sigma '}^{\dagger}
d_{i\gamma\sigma '}d_{i\delta\sigma} \ ,
\end{eqnarray}
where $d_{i\alpha\sigma}^{\dagger}$ creates a hole in the crystal
field state $\alpha$ at site $i$ with spin $\sigma$.
Here the first term is the crystal field Hamiltonian.  The
second term is the spin--orbit interaction,
where $\lambda $ is the spin--orbit coupling constant and
${\bf L}_{\alpha\beta}$ denotes the matrix elements of the
orbital angular momentum vector between the crystal field
states $\alpha$ and $\beta$.  The last term is the
Coulomb interaction, where
\begin{mathletters}
\label{COULOMB}
\begin{eqnarray}
U_{\alpha \beta \gamma \delta} & = & \int d {\bf r}_1 \int
d {\bf r}_2 \psi_\alpha ( {\bf r}_1) \psi_\beta ( {\bf r}_2)
{e^2 \over r_{12} } \psi_\gamma ( {\bf r}_2) \psi_\delta ( {\bf r}_1)
\\ & \equiv & (\alpha \delta | \beta \gamma ) \ .
\end{eqnarray}
\end{mathletters}

\noindent
in the notation \cite{NOTE} of Table A26 of Ref. \onlinecite{GRIFF}.
The second term in Eq. (\ref{MICRO}) gives the Hamiltonian of the
oxygen ions. We assume that the spin--orbit interaction on the
oxygen is much smaller than that on the copper and may be neglected.
Hence we write
\begin{eqnarray}
& &{\cal H}_{\rm O}=\sum_{kn\sigma}\epsilon_{n}p_{kn\sigma}^{\dagger}
p_{kn\sigma}\nonumber\\ &+&\frac{1}{2}
\sum_{n_{1}n_{2}n_{3}n_{4}}\sum_{\sigma\sigma ' k}
U_{n_{1}n_{2}n_{3}n_{4}}
p_{kn_{1}\sigma}^{\dagger}p_{kn_{2}\sigma '}^{\dagger}
p_{kn_{3}\sigma '}p_{kn_{4}\sigma},\label{5}
\end{eqnarray}
in which $p_{kn\sigma}^{\dagger}$ creates a hole in one of the three
$p$-orbitals (denoted by $n$) on the oxygen at site $k$. Here the
Coulomb matrix element is obtained analogously to Eqs. (\ref{COULOMB}).
Finally, ${\cal H}_{\rm hop}$ describes the
hopping between two neighboring oxygen and copper ions:
\begin{equation}
\label{HOP}
{\cal H}_{\rm hop}=\sum_{i\alpha\sigma}\sum_{kn\sigma}t_{\alpha n}^{ik}
d_{i\alpha\sigma}^{\dagger}p_{kn\sigma}+ {\rm hc} \ ,
\end{equation}
in which $t_{\alpha n}^{ik}$ is the hopping matrix element
and hc denotes the Hermitian conjugate of the preceding terms.
As mentioned, it is ${\cal H}_{\rm hop}$ that lifts the
degeneracy of the $2^{N}$-fold degenerate ground state.

In order to derive the effective magnetic Hamiltonian it is
convenient to start from the unperturbed Hamiltonian which
contains the spin-orbit interaction exactly. To achieve
this, we introduce the unitary transformation which diagonalizes
the single--particle part of ${\cal H}_{\rm Cu}$.  As noted
previously, \cite{PRL,PRB} the spin-dependence of
this transformation is fixed by tetragonal symmetry, so
that we can write
\begin{equation}
\label{UNITARY}
d_{i\alpha\sigma}=\sum_{a\sigma '}
m_{\alpha a} [(\sigma (\alpha )]_{\sigma\sigma '}
c_{ia\sigma '},
\end{equation}
where $c_{i a \sigma'}$ destroys a hole in the exact eigenstate
of the Hamiltonian which consists of the crystal field and
spin--orbit interactions.  These states have a site label $i$, a
state label (for which we use roman letters), and a pseudo--spin
index $\sigma'$.  In Eq. (\ref{UNITARY}) we define $\sigma(\alpha)$
for each crystal field state $|\, \alpha \rangle$ as follows:
$\sigma(\alpha) = \sigma_\alpha$ is the Pauli matrix
for $\alpha = x,y,z$ and
$\sigma(\alpha)={\cal I}$ is the unit matrix, for $\alpha=0,1$.
The 5 $\times$ 5 matrix $\bf m$ is the solution to
\begin{equation}
\label{MEQ}
\epsilon_{\alpha} m_{\alpha b} + \sum_\beta {\cal L}_{\alpha\beta}
m_{\beta b} = E_{b} m_{\alpha b}
\end{equation}
with $\sum_\alpha m^\ast_{\alpha a} m_{\alpha b} = \delta_{ab}$,
where $\delta$ is the Kronecker delta function.  Here
${\cal L}_{\alpha \beta}$ are related to the matrix elements
of the orbital angular momentum vector and are given by
\begin{eqnarray}
& &{\cal L}_{0z}=-i\lambda,\ \
{\cal L}_{0x}={\cal L}_{0y}=i\lambda /2,\nonumber\\
& & {\cal L}_{1x}=-{\cal L}_{1y}=i\sqrt{3}\lambda /2,\nonumber\\
& &{\cal L}_{zx}={\cal L}_{zy}={\cal L}_{xy}=\lambda /2. \label{10b}
\end{eqnarray}
Matrix elements not listed and not obtainable using
${\cal L}_{\alpha \beta} = {\cal L}_{\beta \alpha}^\ast$
are zero.  When $\lambda \rightarrow 0$, each state $|\, a \rangle$
approaches one of the states $|\, \alpha \rangle$.  Using this
identification, the indices $a$ also run over the values
$0,1,z,x$, and $y$.  We use the values of the parameters
which are listed in Table I and are discussed in detail in YHAE.

\section{PERTURBATION EXPANSION}

We now divide the total Hamiltonian ${\cal H}$ into an unperturbed
part, ${\cal H}_{0}$, and a perturbation term ${\cal H}_{1}$. The
part ${\cal H}_{0}$ contains the single--particle Hamiltonians on
the coppers and on the oxygens, {\it and } the leading on--site
Coulomb potentials,
\begin{eqnarray}
\label{H0EQ}
& &{\cal H}_{0}=\sum_{ia\sigma}E_{a}c_{ia\sigma}^{\dagger}
c_{ia\sigma}+\frac{U_{0}}{2}
\sum_{\stackrel{iab}{\sigma\sigma '}}
c_{ia\sigma}^{\dagger}c_{ib\sigma '}^{\dagger}
c_{ib\sigma '}c_{ia\sigma}\nonumber\\
& &+\sum_{kn\sigma}\epsilon_{n}p_{kn\sigma}^{\dagger}
p_{kn\sigma}+\frac{U_{P}}{2}
\sum_{\stackrel{knn'}{\sigma\sigma '}}p_{kn\sigma}^{\dagger}
p_{kn'\sigma '}^{\dagger}
p_{kn'\sigma '}p_{kn\sigma}\ .
\end{eqnarray}
For tetragonal site symmetry, we choose
$U_0 \equiv U_{\alpha \alpha \alpha \alpha} = A+4B+C$ and
$U_P \equiv U_{nnnn}$.
The perturbation Hamiltonian includes the kinetic energy and the
remaining Coulomb interactions. Because of the transformation
(\ref{UNITARY}), the hopping becomes spin--dependent. Thus we have
\begin{equation}
{\cal H}_{1}={\cal H}_{\rm hop}+\Delta{\cal H}_{C},\label{12}
\end{equation}
in which
\begin{equation}
{\cal H}_{\rm hop}=\sum_{ia\sigma}\sum_{kn\sigma '}
(\tilde{t}_{an}^{ik})_{\sigma '\sigma}
c_{ia\sigma '}^{\dagger}p_{kn\sigma}+ {\rm hc} , \label{13}
\end{equation}
with the $2\times 2$ matrix
\begin{equation}
\label{TTEQ}
\tilde{t}_{an}^{ik}=\sum_{\alpha}t_{\alpha n}^{ik}
m_{\alpha a}^{\ast}\sigma (\alpha )\ ,
\end{equation}
and $\Delta{\cal H}_{C}$ describes the perturbation parts of the
on--site Coulomb potentials,
\begin{eqnarray}
& &\Delta{\cal H}_{C}=
\frac{1}{2}\sum_{\stackrel{ss's_{1}s_{1}'}{iabcd}}
\Delta\bar{U}_{ss's_{1}s_{1}'}(abcd)
c_{ias}^{\dagger}c_{ibs'}^{\dagger}c_{ics_{1}}c_{ids_{1}'}\nonumber\\
& &+\frac{1}{2}\sum_{\stackrel{k\sigma\sigma '}{n_{1}n_{2}n_{3}n_{4}}}\Delta
U_{n_{1}n_{2}n_{3}n_{4}}p_{kn_{1}\sigma}^{\dagger}p_{kn_{2}\sigma '}^{\dagger}
p_{kn_{3}\sigma '}p_{kn_{4}\sigma},\label{15}
\end{eqnarray}
with
\begin{eqnarray}
& &\Delta \bar{U}_{ss's_{1}s_{1}'}(abcd)=\sum_{\alpha\beta\gamma\delta}\Delta
U_{\alpha\beta\gamma\delta}\nonumber\\
& & \times m_{\alpha a}^{\ast}m_{\beta b}^{\ast}m_{\gamma c}
m_{\delta d}[\sigma (\alpha )\sigma (\delta )]_{ss_{1}'}[\sigma (\beta )
\sigma (\gamma )]_{s's_{1}}.\label{16}
\end{eqnarray}
Note that $\Delta {\bf U}$ involves only the small Racah coefficients,
$B$ and $C$.

The effective magnetic interaction ${\cal H}(i,j)$ between
magnetic ions $i$ and $j$ is found by perturbation expansion
with respect to ${\cal H}_{1}$.  All the perturbation
contributions to ${\cal H}(i,j)$ should involve products of
matrix elements which begin and end within the $2^{N}$--fold
degenerate ground--state manifold of ${\cal H}_{0}$, each state
of which has one hole at each copper site, with
arbitrary spin $\sigma $. Denoting such a ground state by
$\mid\psi_{0}\rangle$, and concentrating on perturbation terms which
involve only two coppers $i$ and $j$ and one oxygen between them,
it is convenient to replace $\mid\psi_{0} \rangle$ by
$\sum_{\sigma\sigma_{1}}c_{i0\sigma}^{\dagger}c_{j0\sigma_{1}}^{\dagger}
c_{j0\sigma_{1}}c_{i0\sigma}\mid\psi_{0} \rangle$. This ensures that
$\mid\psi_{0}\rangle$ indeed has exactly one hole on each copper $i$ and
$j$. Clearly the lowest--order contributions to the energy are of
order $\tilde t^{4}$. There are two possible channels in this order, which
we denote by $a$ and $b$. In channel $a$, the hole is transferred
from one of the coppers to the oxygen, then to the second copper.
Afterwards, one of the holes on this second copper
returns to the empty copper. Hence in
this channel there are two holes on the copper in an intermediate
state. In channel $b$, the hole is transferred from one of the
coppers to the oxygen, and then a second hole is taken from another
copper to the same oxygen. Then the two holes hop back,
each to one of the initial coppers. Thus in channel $b$ there are two
holes on the oxygen in an intermediate state. When the perturbation
contributions coming from the Coulomb potential $\Delta{\cal H}_{C}$
[Eqs. (\ref{12}), (\ref{15}) and (\ref{16})] are included, the
on--site interactions on the copper are effective in channel $a$,
and those on the oxygen appear in channel $b$.
These contributions are of order $\tilde t^{4}\Delta{\cal H}_{C}$.

The perturbation contributions to order $\tilde t^{4}$ are the same
as those given by YHAE and are
\begin{equation}
\label{H1T4}
{\cal H}^{(1)}(i,j)=\sum_{nn'}g_{nn'} {\rm Tr} \Bigl\{\vec{\sigma}
\cdot {\bf S}_{i}\tilde{t}_{0n}^{ik}\tilde{t}_{n0}^{kj}
\vec{\sigma}\cdot{\bf S}_{j}\tilde{t}_{0n'}^{jk}
\tilde{t}_{n'0}^{ki}\Bigr\}\ ,
\end{equation}
in which
\begin{equation}
{\bf S}_{i}=\frac{1}{2}\sum_{\sigma\sigma '}c_{i0\sigma}^{\dagger}
(\vec{\sigma})_{\sigma\sigma '}c_{i0\sigma '},\label{18}
\end{equation}
is the spin on the copper at site $i$ in the orbital state
$| \, 0 \rangle$, $\tilde{t}_{0n}^{ik}$ is the $2\times 2$
matrix given by Eq.  (\ref{TTEQ}) and
\begin{equation}
g_{nn'}=\frac{2}{U_{0}}\frac{1}{\epsilon_{n}\epsilon_{n'}}+
\frac{1}{U_{P}+\epsilon_{n}+\epsilon_{n'}}\Bigl(\frac{1}
{\epsilon_{n}}+\frac{1}{\epsilon_{n'}}\Bigr)^{2}.\label{19}
\end{equation}
The first term in $g_{nn'}$ results from channel $a$ while the second arises
from channel $b$. In deriving expression (\ref{H1T4}) we have assumed that
the single-particle energy $E_{a=0}$ on the copper is equal to zero.

 From the form of ${\cal H}^{(1)}(i,j)$ it is clear that it may yield
anisotropic magnetic interactions only when the
effective hopping between the coppers, $\tilde{t}^{ik}\tilde{t}^{kj}$,
involves spin-flips. When the hopping conserves the spin,
the $\tilde{t}$ products are proportional to the
$2\times 2$ unit matrix and the trace in Eq. (\ref{H1T4})
gives a result proportional to ${\bf S}_{i}\cdot {\bf S}_{j}$.
This is indeed the case for
tetragonal symmetry, as we discuss in the next section. \cite{PRL,PRB}
In the orthorhombic phase of La$_{2}$CuO$_{4}$,
however, the effective hopping between the coppers is accompanied
by spin-flip, and consequently ${\cal H}^{(1)}(i,j)$ includes the
antisymmetric Dzyaloshinskii-Moriya interaction, as well as symmetric
magnetic anisotropies. \cite{ANIS3,ANIS2}

We now turn to the contributions of the Coulomb potential.  In particular,
we consider contributions to the magnetic energy which are of order
$\tilde t^4$ and first order in $\Delta{\cal H}_{C}$.  Following
the approach of YHAE one finds that the processes
from channel $a$ (indicated by a subscript "$a$") yield
\begin{eqnarray}
\label{20}
& &{\cal H}^{(2)}_a(i,j)= \Biggl[ \sum_{\stackrel{a,b,\neq0}
{\alpha\beta\gamma\delta}}
\frac{\Delta U_{\alpha\beta\gamma\delta}}{(U_{0}+E_{a})
(U_{0}+E_{b})}\nonumber\\
& & \times \Biggl( {\rm Tr} \Bigl\{\vec{\sigma}\cdot {\bf S}_{i}
\sum_{n}\frac{\tilde{t}_{0n}^{ik}
\tilde{t}_{nb}^{kj}}{\epsilon_{n}}\sigma (\alpha )\sigma (\delta )
\sum_{n'}\frac{\tilde{t}_{an'}^{jk}\tilde{t}_{n'0}^{ki}}
{\epsilon_{n'}}\Bigr\}\nonumber\\
& & \times  {\rm Tr} \Bigl\{\vec{\sigma}\cdot {\bf S}_{j}
\sigma (\beta )\sigma (\gamma )\Bigr\}
m_{\alpha b}^{\ast}m_{\beta 0}^{\ast}m_{\gamma 0}m_{\delta a}\nonumber\\
& &- {\rm Tr} \Bigr\{\vec{\sigma}
\cdot {\bf S}_{i}\sum_{n}\frac{\tilde{t}_{0n}^{ik}
\tilde{t}_{nb}^{kj}}{\epsilon_{n}}\sigma
(\alpha )\sigma (\delta )\vec{\sigma}\cdot {\bf S}_{j}
\sigma (\beta )\sigma (\gamma )\nonumber\\
& & \times \sum_{n'}\frac{\tilde{t}_{an'}^{jk}\tilde{t}_{n'0}^{ki}}
{\epsilon_{n'}}\Bigr\}
m_{\alpha b}^{\ast}m_{\beta 0}^{\ast}m_{\gamma a}m_{\delta 0}\Biggr)
+(i\leftrightarrow j) \Biggr] ,
\end{eqnarray}
where $i \leftrightarrow j$ denotes the sum of previous terms
with $i$ and $j$ interchanged.
Here it is convenient to classify the Coulomb matrix elements into
two classes, by writing
\begin{equation}
\Delta U_{\alpha \beta \gamma \delta} = U^{(1)}_{\alpha \beta \gamma \delta}
+ U^{(2)}_{\alpha \beta \gamma \delta}\ ,
\end{equation}
where $U^{(1)}_{\alpha \beta \gamma \delta}$ is nonzero only if
\begin{mathletters}
\begin{equation}
\label{U1COND}
\sigma(\alpha)= \sigma(\delta) \ , \ \ {\rm and}
\ \ \sigma(\beta)=\sigma(\gamma) \ ,
\end{equation}
while $U^{(2)}_{\alpha \beta \gamma \delta}$ is nonzero only if
\begin{equation}
\label{U2COND}
\sigma(\alpha) \sigma(\delta) = C_1 \sigma_\mu = C_2
\sigma(\beta) \sigma(\gamma) \ , \ \ {\rm for} \ \ \mu=x,y,z,
\end{equation}
\end{mathletters}

\noindent
where $C_1$ and $C_2$ are constants which may be imaginary.  This
classification will be of immediate use.  For the elements denoted
$U_{\alpha\beta\gamma\delta}^{(1)}$, the products
$\sigma (\alpha )\sigma (\delta )$ and $\sigma (\beta )\sigma (\gamma )$
are both proportional to the unit matrix. Therefore the first term in the
square brackets of Eq. (\ref{20}) vanishes, and one is left with the second
term alone. For the matrix elements denoted by
$U_{\alpha\beta\gamma\delta}^{(2)}$,
both products $\sigma (\alpha )\sigma (\delta )$ and
$\sigma (\beta )\sigma (\gamma )$ are proportional to
$\sigma_{\mu}$, $\mu =x,y,z$.
In treating this contribution it is convenient to use the identity
\begin{equation}
\sigma_{\mu}\vec{\sigma}\cdot {\bf S}\sigma_{\mu}=2\sigma_{\mu}S_{\mu}
-\vec{\sigma}\cdot {\bf S}.\label{21}
\end{equation}
We see that the matrix elements
$U_{\alpha\beta\gamma\delta}^{(2)}$ give rise to
magnetic interactions which depend on the Cartesian index $\mu $.
Rearranging the terms in (\ref{20}) and defining
\begin{mathletters}
\begin{eqnarray}
\label{22a}
& &Q_{ab}=\sum_{\alpha\beta\gamma\delta}
\frac{U_{\alpha\beta\gamma\delta}^{(2)}
- U_{\alpha\beta\gamma\delta}^{(1)}}
{(U_{0}+E_{a})(U_{0}+E_{b})}m_{\alpha b}^{\ast}
m_{\beta 0}^{\ast}m_{\gamma 0}m_{\delta a}, \\
\label{22b}
& &K_{ab}^{\mu}=
\sum_{\alpha \beta \gamma \delta}
%\sum_{\stackrel{\alpha\delta}{\sigma
%(\alpha )\sigma (\delta
%)\sim\sigma_{\mu}}}\sum_{\stackrel{\beta\gamma}{\sigma
%(\beta)\sigma (\gamma )\sim\sigma_{\mu}}}
\frac{U_{\alpha\beta\gamma\delta}^{(2)}}{(U_{0}+E_{a})
(U_{0}+E_{b})}\nonumber\\
& & \times (m_{\alpha b}^{\ast}m_{\beta 0}^{\ast}-m_{\beta b}^{\ast}
m_{\alpha 0}^{\ast}) (m_{\delta a}m_{\gamma 0}-m_{\delta 0}m_{\gamma a}),
\end{eqnarray}
\end{mathletters}
we obtain
\begin{eqnarray}
\label{23}
& &{\cal H}^{(2)}_a (i,j)=\nonumber\\
& &\sum_{a,b\neq 0}\Biggl[Q_{ab} {\rm Tr} \Bigl\{\vec{\sigma}\cdot {\bf S}_{i}
\sum_{n}\frac{\tilde{t}_{0n}^{ik}\tilde{t}_{nb}^{kj}}{\epsilon_{n}}
\vec{\sigma}\cdot {\bf S}_{j}\sum_{n'}\frac{\tilde{t}_{an'}^{jk}
\tilde{t}_{n'0}^{ki}}{\epsilon_{n'}}\Bigr\}\nonumber\\
& &+\sum_{\mu}K_{ab}^{\mu}S_{j\mu}
{\rm Tr} \Bigr\{\vec{\sigma}\cdot {\bf S}_{i}\sum_{n}
\frac{\tilde{t}_{0n}^{ik}\tilde{t}_{nb}^{kj}}{\epsilon_{n}}\sigma_{\mu}
\sum_{n'}\frac{\tilde{t}_{an'}^{jk}
\tilde{t}_{n'0}^{ki}}{\epsilon_{n'}}\Bigr\} \nonumber \\
&& + (i\leftrightarrow j)\Biggr].
\end{eqnarray}
Note that the sum over state labels in Eqs. (\ref{22a}) and (\ref{22b})
is restricted by the conditions of Eqs. (\ref{U1COND}) and (\ref{U2COND}).
One notes that the first term in Eq. (\ref{23})
has a structure similar to that of
${\cal H}^{(1)}(i,j)$ in Eq. (\ref{H1T4}) and therefore has the same magnetic
symmetries. The second term, however, leads to magnetic anisotropy even for
tetragonal symmetry, for which the effective hopping between the coppers is
spin-independent. This is elaborated upon in the next section.

Finally we consider the processes in channel $b$, in which the two
holes are on the oxygen in an intermediate state and thus experience the
Coulomb interaction on the oxygen. These processes yield
\begin{eqnarray}
\label{24}
& &{\cal H}^{(2)}_b (i,j) = \sum_{nn'n_{1}n_{1}'}(\frac{1}{\epsilon_{n}}
+ \frac{1}{\epsilon_{n'}})(\frac{1}{\epsilon_{n_{1}}}
+ \frac{1}{\epsilon_{n_{1}'}})\nonumber\\
&\times &\frac{\Delta U_{n_{1}n_{1}'nn'}}{(U_{P}+\epsilon_{n}+\epsilon_{n'})
(U_{P}+\epsilon_{n_{1}}+\epsilon_{n_{1}'})}\nonumber\\
& \times  & \Biggl[ {\rm Tr}
\Bigr\{\vec{\sigma}\cdot {\bf S}_{i}\tilde{t}_{0n_{1}}^{ik}
\tilde{t}_{n' 0}^{ki}\Bigr\} {\rm Tr} \Bigl\{\vec{\sigma}\cdot {\bf S}_{j}
\tilde{t}_{0n_{1}'}^{jk}\tilde{t}_{n0}^{kj}\Bigr\}\nonumber\\
&-& {\rm Tr} \Bigr\{\vec{\sigma}\cdot {\bf S}_{i}
\tilde{t}_{0n_{1}}^{ik}
\tilde{t}_{n' 0}^{kj}\vec{\sigma}\cdot {\bf S}_{j}
\tilde{t}_{0n_{1}'}^{jk}\tilde{t}_{n0}^{ki}\Bigr\}\Biggr]\ .
\end{eqnarray}

To the order we work, the total magnetic interaction between a pair of
copper ions is given by
\begin{equation}
\label{25}
{\cal H}(i,j)={\cal H}^{(1)}(i,j)+{\cal H}^{(2)}_a(i,j)
+{\cal H}^{(2)}_b(i,j)\ .
\end{equation}
In the next section we examine ${\cal H}(i,j)$ in the tetragonal phase.
In the orthorhombic phase the situation is more complicated.  In
principle, one should start by replacing the tetragonal crystal
field states by those which are calculated in the presence of the
orthorhombic distortion.  A reasonable approximation is to take
account of this distortion by considering the crystal field
states of the octahedron of oxygen ions assuming this octahedron
to be rotated rigidly away from its orientation in the tetragonal
phase.  Then, to the extent that the crystal field is wholly determined
by the octahedron of oxygen neighbors, it will be tetragonal in
the rotated coordinate system fixed by the shell of oxygen
neighbors.  Then the result of Eq. (\ref{25}) can be used.
The major complication is that the hopping matrix elements
are those between rotated orbitals, as will be detailed
elsewhere.\cite{STEIN}

Some comments concerning the applicability of our results
should be made.  As discussed in YHAE, we believe that
the use of perturbation theory is justified for the
calculation of the anisotropy of the superexchange Hamiltonian.
In contrast, for the isotropic terms it has been shown\cite{JHJ}
that perturbation theory is not reliable, mainly because
the energies of the excited states of the oxygen
levels are not very large in comparison to $t$.  However, since
the spin--orbit interaction takes place on the copper ions, this
effect is much reduced in the anisotropic terms.  In addition,
there are contributions to the isotropic interactions which
can not be obtained by consideration of a single bond.
In one of these, a hole hops from a copper to one nearest--neighboring
oxygen, then diagonally to a different nearest--neighbor oxygen, and
finally back to the original copper.\cite{JHJ}  Also, the
effective spin Hamiltonian acquires isotropic contributions
from processes of order $\tilde t^8$ which involve a hole hopping
around a plaquette of Cu ions.\cite{TAKA}  These plaquette
interactions involve both two--spin interactions (between nearest
and next--nearest neighbors) and four--spin interactions.

\section{Tetragonal symmetry}

Here we study the effective magnetic Hamiltonian ${\cal H}(i,j)$
[Eqs. (\ref{H1T4}) and (\ref{23})-(\ref{25})] for the case of
tetragonal symmetry. For this symmetry the effective
interaction which is bilinear in the spin operators must be
of the form
\begin{equation}
\label{27}
{\cal H} (i,j) = \sum_\mu J_{\mu \mu} (i,j) S_\mu (i)
S_\mu (j) \ ,
\end{equation}
where $\mu = x,y,z$.  Since we are interested in the
{\it anisotropic} part of the exchange interaction, we will
drop any contributions which we identify as being isotropic.
Obviously, our results can not be used for the magnitude of
an individual $J_{\mu \mu}$, but rather apply to the
difference between two such quantities.

Investigation of Eqs. (\ref{H1T4}), (\ref{23}) and (\ref{24})
reveals that one can define an effective hopping between the
copper ions, which is generally given by the product
$\tilde{t}_{an}^{ik}\tilde{t}_{n'b}^{kj}$. (The interaction
${\cal H}^{(2)}_b (i,j)$ [Eq. (\ref{24})] requires also the
cases $i=j$.) We therefore start by examining this quantity.
To this end we note that the nonzero hopping matrix elements
between the tetragonal states $\mid\alpha \rangle$ on the coppers and
the states $\mid n \rangle$ on the oxygen are
$t_{0p_{x}}$, $t_{1p_{x}}$, $t_{yp_{z}}$ and $t_{zp_{y}}$ for
a bond along the $x$-direction in the CuO$_{2}$ plane, and
analogously $t_{0p_{y}}$, $t_{1p_{y}}$, $t_{xp_{z}}$ and
$t_{zp_{x}}$ for a bond along $y$.\cite{PRL,PRB}
Using now Eq. (\ref{UNITARY}) we find
\begin{equation}
\tilde{t}_{an}\tilde{t}_{n'b}=\sum_{\alpha\beta}t_{\alpha n}t_{n'\beta}
m_{\alpha a}^{\ast}m_{\beta b}\sigma (\alpha )\sigma (\beta ),\label{26}
\end{equation}
where we have omitted the site indices for convenience.
It therefore follows that in the case $n=n'$ the quantity in
Eq. (\ref{26}) is proportional to the $2\times 2$ unit matrix,
namely, the effective hopping between the copper ions is
{\it not} accompanied by spin-flip. This means that in the
expressions for the interactions ${\cal H}^{(1)}(i,j)$ and
${\cal H}^{(2)}_a(i,j)$ [Eqs. (\ref{H1T4}) and (\ref{23})]
we may take all $\tilde{t}$'s outside the trace. Consequently,
the contribution to the magnetic energy from Eq. (\ref{H1T4})
is isotropic, and so is that from the first term in Eq. (\ref{23}),
proportional to $Q_{ab}$. The second term in ${\cal H}^{(2)}_a(i,j)$,
which arises from
$U_{\alpha\beta\gamma\delta}^{(2)}$, leads to an anisotropic
magnetic interaction of the form of Eq. (\ref{27}), with
\begin{equation}
J_{\mu\mu}=4\sum_{a,b\neq 0}K_{ab}^{\mu}\sum_{n}
\frac{\tilde{t}_{0n}\tilde{t}_{nb}}
{\epsilon_{n}}\sum_{n'}\frac{\tilde{t}_{an'}\tilde{t}_{n'0}}
{\epsilon_{n'}}.\label{28}
\end{equation}

We now consider the magnetic symmetry of ${\cal H}^{(2)}_b(i,j)$,
which results from the on--site Coulomb potential on the oxygen.  The
only nonzero matrix elements $\Delta U_{n_{1}n_{1}'nn'}$ are \cite{UOX}
\begin{equation}
\Delta U_{nn'n'n},\ \ \Delta U_{nn'nn'},\ \ \Delta U_{nnn'n'}.\label{29}
\end{equation}
The first of these implies terms of the form $\tilde{t}_{0n}\tilde{t}_{n0}$
[{\it cf. }Eq. (\ref{24})], which are proportional to the unit matrix. As a
result the first term in the square brackets of (\ref{24}) disappears. The
second, which yields an isotropic interaction, can be combined into
${\cal H}^{(1)}(i,j)$ by redefining $g_{nn'}$ [Eq. (\ref{19})] to be
\begin{eqnarray}
& &g_{nn'}=\frac{2}{U_{0}}\frac{1}{\epsilon_{n}\epsilon_{n'}}\nonumber\\
&+&\frac{1}{U_{P}+\Delta U_{nn'n'n}+\epsilon_{n}+\epsilon_{n'}}
\Bigl(\frac{1}{\epsilon_{n}}+\frac{1}{\epsilon_{n'}}\Bigr)^{2}.\label{30}
\end{eqnarray}
For simplicity we have put the $\Delta U$ in the denominator of
this expression.  The results of this paper are correct to first
order in $\Delta {\cal H}_c$.

The other two matrix elements of Eq. (\ref{29}) lead to
\begin{eqnarray}
& &{\cal H}^{(2)}_b(i,j)=\nonumber\\
& &\sum_{\stackrel{nn'}{n\neq n'}}\
\Biggl[\Delta
U_{nn'nn'}\Bigl(\frac{1}{\epsilon_{n}}+\frac{1}{\epsilon_{n'}}\Bigr)^{2}
\Bigl(\frac{1}{U_{P}+\epsilon_{n}+\epsilon_{n'}}\Bigr)^{2}\nonumber\\
&  \times &\Bigl({\rm Tr}\Bigl\{\vec{\sigma}\cdot {\bf S}_{i}\tilde{t}_{0n}
\tilde{t}_{n'0}\Bigr\} {\rm Tr}
\Bigl\{\vec{\sigma}\cdot {\bf S}_{j}\tilde{t}_{0n'}
\tilde{t}_{n0}\Bigr\}\nonumber\\
&-&{\rm Tr} \Bigl\{\vec{\sigma}\cdot{ \bf S}_{i}\tilde{t}_{0n}\tilde{t}_{n'0}
\vec{\sigma}\cdot{ \bf S}_{j}\tilde{t}_{0n'}
\tilde{t}_{n0}\Bigr\}\Bigr)\nonumber\\
&+&\Delta U_{nn'nn'}\frac{4}{\epsilon_{n}\epsilon_{n'}}
\frac{1}{(U_{P}+2\epsilon_{n})(U_{P}+2\epsilon_{n'})}\nonumber\\
&  \times &\Bigl( {\rm Tr} \Bigl\{\vec{\sigma}\cdot{ \bf S}_{i}\tilde{t}_{0n}
\tilde{t}_{n'0}\Bigr\}
{\rm Tr}\Bigl\{\vec{\sigma}\cdot{ \bf S}_{j}\tilde{t}_{0n}
\tilde{t}_{n'0}\Bigr\}\nonumber\\
&-&{\rm Tr} \Bigl\{\vec{\sigma}\cdot{ \bf S}_{i}\tilde{t}_{0n}\tilde{t}_{n'0}
\vec{\sigma}\cdot{ \bf S}_{j}\tilde{t}_{0n}\tilde{t}_{n'0}
\Bigr\}\Bigr)\Biggr].\label{31}
\end{eqnarray}
The main point to note here is that although
$\tilde{t}_{0n}\tilde{t}_{n'0}$ implies a spin--flip in the sense
that this product is proportional to one of the Pauli matrices
[{\it cf.} Eq. (\ref{26})], it is the {\it same} Pauli matrix
for both products appearing in each term of Eq. (\ref{31}).
This makes the resulting interaction isotropic. For example,
writing $\tilde{t}_{0n}\tilde{t}_{n'0}\propto\sigma_{\mu}$,
with $\mu =x,y$ or $z$, we find that the spin dependence in each
of the terms in Eq. (\ref{31}) is proportional to
\begin{eqnarray}
& &{ \rm Tr} \Bigl\{\vec{\sigma}\cdot{ \bf S}_{i}\sigma_{\mu}\Bigr\}
{\rm Tr} \Bigl\{\vec{\sigma}\cdot{ \bf S}_{j}\sigma_{\mu}\Bigr\}\nonumber\\
&-&{\rm Tr} \Bigl\{\vec{\sigma}\cdot {\bf S}_{i}\sigma_{\mu}
\vec{\sigma}\cdot {\bf S}_{j}\sigma_{\mu}\Bigr\}
=2{\bf S}_{i}\cdot {\bf S}_{j},\label{32}
\end{eqnarray}
where we have used the identity (\ref{21}). Thus the perturbation
processes for which there are two holes on the oxygen in the
intermediate state do not contribute to the magnetic anisotropy
in tetragonal symmetry. This result holds because the spin-orbit
interaction on the oxygen has been discarded.

The anisotropic exchange interactions in the tetragonal phase
thus come exclusively from channel $a$.  Also, as previously noted,
only the term proportional to ${\bf K}$ in Eq. (\ref{23}) gives
rise to anisotropy.  Furthermore one notes that there is no
anisotropic contribution to $J_{\mu\mu}$ from the matrix elements
$U_{\alpha\beta\gamma\delta}^{(2)}$ for which $\alpha =\beta$
or $\gamma =\delta$.

Up to now, our results have been valid to all orders in the
spin--orbit coupling, $\lambda $. We now obtain an explicit
expression for the anisotropic part of $J_{\mu \mu}$ to leading
order in $\lambda $, which turns out to be ${\cal O}(\lambda ^{2})$.
To this end we use Eq. (\ref{MEQ}), from which we get
\begin{equation}
\label{33}
m_{\alpha a}= \left\{
\begin{array} {l l}
1 & {\rm for} \ \alpha = a \\
\frac{{\cal L}_{\alpha a}}
{\epsilon_{a}-\epsilon_{\alpha}} \ \
& {\rm for} \ \alpha \not= a
\end{array} \ + {\cal O}(\lambda^{2}),
\right.
\end{equation}
where ${\cal L}_{\alpha a}$ for $\alpha\neq a$ [see Eqs. (\ref{10b})]
is of order $\lambda$. Next we note that the summation indices $a$
and $b$ of Eq. (\ref{28}) can take the values $\mid a\rangle,
\mid b \rangle =\mid 1 \rangle$
or $\mid a \rangle , \mid b \rangle =\mid x\rangle ,\mid y \rangle$
or $\mid z \rangle$. We therefore need to evaluate
$\sum_{n}\tilde{t}_{0n}\tilde{t}_{nb}/\epsilon_{n}$
for $\mid b \rangle =\mid 1 \rangle$ and for
$\mid b \rangle =\mid \nu \rangle$, where $\nu =x,y,z.$
Using Eqs. (\ref{TTEQ}) and (\ref{33}) we find
\begin{eqnarray}
& &\sum_{n}\frac{\tilde{t}_{0n}\tilde{t}_{n1}}{\epsilon_{n}}=\sum_{n}
\frac{t_{0n}t_{n1}}{\epsilon_{n}}+{\cal O}(\lambda^{2}),\nonumber\\
& &\sum_{n}\frac{\tilde{t}_{0n}\tilde{t}_{n\mu}}{\epsilon_{n}}=\sum_{n}
\frac{{t}_{0n}^{2}-t_{\mu n}^{2}}{\epsilon_{n}}\frac{{\cal L}_{0\mu}}
{\epsilon_{\mu}}\nonumber\\
&+&\sum_{n}\frac{t_{0n}t_{n1}}{\epsilon_{n}}
\frac{{\cal L}_{1\mu}}{\epsilon_{\mu}-\epsilon_{1}}+{\cal
O}(\lambda^{2}),\label{34}
\end{eqnarray}
where the $t$'s are the hopping matrix elements for the tetragonal
states. Equations (\ref{33}) and (\ref{34}), in conjunction with
Eq. (\ref{22b}), imply that the contributions of
$U_{\alpha\beta\gamma\delta}^{(2)}$ where more than two of the
indices $\alpha$, $\beta $, $\gamma$, and $\delta$ take the values
$x$, $y$ or $z$ are at least of order $\lambda^{3}$. Take for
example the element $U_{1zxy}^{(2)}$.
 From Eqs. (\ref{22b}) and (\ref{33}) we see that the $m$-products
in (\ref{22b}) are of order $\lambda^{2}$ provided that
$\mid b \rangle =\mid 1 \rangle$ and $\mid a \rangle =\mid x \rangle$
or $\mid y \rangle$. In all other cases the $m$-products are at
least of order $\lambda^{3}$. But with this choice [see Eqs. (\ref{34})]
the hopping matrix elements provide another factor of $\lambda$,
to render $J_{\mu\mu}$ to be of order $\lambda^{3}$. It follows
that out of all nonzero $U_{\alpha\beta\gamma\delta}^{(2)}$, the
ones that contribute to $J_{\mu\mu}$ up to order $\lambda^{2}$ are
\begin{eqnarray}
& &U_{0\mu 0\mu}^{(2)}=U_{\mu 0\mu 0}^{(2)},\ \ U_{1\mu 1\mu}^{(2)}
=U_{\mu 1\mu 1}^{(2)},\nonumber\\
& &U_{0\mu 1\mu}^{(2)}=U_{\mu 0\mu 1}^{(2)}=U_{\mu 1\mu 0}^{(2)}
=U_{1\mu 0\mu}^{(2)}.\label{35}
\end{eqnarray}
We are now in position to calculate $K_{ab}^{\mu}$. When both
$\mid a\rangle$ and $\mid b\rangle$ are equal to $\mid 1\rangle$
one finds
\begin{eqnarray}
\label{36}
& &K_{11}^{\mu}=\frac{2}{(U_{0}+\epsilon_{1})^{2}}\Biggl[-
\frac{{\cal L}_{\mu 1}{\cal L}_{1\mu}}{(\epsilon_{1}-\epsilon_{\mu})^{2}}
U_{0\mu 0\mu}^{(2)}\nonumber\\
&-&\frac{{\cal L}_{\mu 0}{\cal L}_{0\mu}}{\epsilon_{\mu}^{2}}U_{1\mu
1\mu}^{(2)}
-\frac{{\cal L}_{1\mu}{\cal L}_{\mu 0}+{\cal L}_{0\mu}{\cal L}_{\mu 1}}
{\epsilon_{\mu}(\epsilon_{1}-\epsilon_{\mu})}U_{0\mu 1\mu}^{(2)}\Biggr] \ .
\end{eqnarray}
When $\mid a\rangle =\mid 1\rangle$ and $\mid b\rangle =\mid\mu \rangle$
or {\it vice versa}, we have
\begin{eqnarray}
& &K_{1\mu}^{\mu}=(K_{\mu 1}^{\mu})^{\ast}
=\frac{2}{(U_{0}+\epsilon_{1})(U_{0}+\epsilon_{\mu})}\nonumber\\
& \times  &\Biggl[-
\frac{{\cal L}_{\mu 1}}{\epsilon_{1}-\epsilon_{\mu}}U_{0\mu 0\mu}^{(2)}
-\frac{{\cal L}_{\mu 0}}{\epsilon_{\mu}}U_{0\mu 1\mu}^{(2)}\Biggr],\label{37}
\end{eqnarray}
where we have kept terms up to order $\lambda $ since the hopping
elements in this case will contribute another factor of $\lambda $.
Finally, the case where both $\mid a\rangle$ and $\mid b\rangle$
are equal to $\mid\mu \rangle$ requires terms to order $\lambda^{0}$
and is therefore
\begin{equation}
\label{38}
K_{\mu\mu}^{\mu}=-\frac{2}{(U_{0}+\epsilon_{\mu})^{2}} U_{0\mu 0\mu}^{(2)}.
\end{equation}

Combining the results of Eqs. (\ref{34})-(\ref{38}) we obtain
\begin{eqnarray}
& &J_{\mu\mu}=-\frac{8U_{0\mu 0\mu}^{(2)}}{(U_{0}+\epsilon_{\mu})^{2}}
\Biggl|
\frac{{\cal L}_{\mu 1}}{U_{0}+\epsilon_{1}}\sum_{n}\frac{t_{0n}t_{n1}}
{\epsilon_{n}}
\nonumber\\
&-&\frac{{\cal L}_{\mu 0}}{\epsilon_{\mu}}\sum_{n}\frac{t_{0n}^{2}-t_{\mu
n}^{2}}{\epsilon_{n}}
\Biggr|^{2}
-\frac{8U_{1\mu 1\mu}^{(2)}}{(U_{0}+\epsilon_{1})^{2}}
\Biggl|\frac{{\cal L}_{\mu 0}}{\epsilon_{\mu}}
\sum_{n}\frac{t_{0n}t_{n1}}{\epsilon_{n}} \Biggr|^{2}\nonumber\\
&-&\frac{16U_{0\mu 1\mu}^{(2)}}{(U_{0}+\epsilon_{1})(U_{0}+\epsilon_{\mu})}
\frac{{\cal L}_{\mu 0}}{\epsilon_{\mu}}
\sum_{n}\frac{t_{0n}t_{n1}}{\epsilon_{n}}\nonumber\\
& \times  &\Bigl[\frac{{\cal L}_{0\mu}}{\epsilon_{\mu}}
\sum_{n}\frac{t_{0n}^{2}-t_{\mu n}^{2}}
{\epsilon_{n}}
-\frac{{\cal L}_{1\mu}}{U_{0}+\epsilon_{1}}\sum_{n}\frac{t_{0n}t_{n1}}
{\epsilon_{n}}\Bigr].\label{39}
\end{eqnarray}
It remains to insert here the explicit expressions for the hopping terms,
Eqs. (\ref{34}), in conjunction with the expressions for the angular momentum
matrix elements, ${\cal L}_{\alpha\beta}$, Eqs. (\ref{10b}). To be specific,
we consider a bond along $x$, and use the values of
$U_{\alpha \beta \gamma \delta} \equiv (\alpha \delta | \, \beta \gamma )$
as listed in Table II.  One then finds
\begin{mathletters}\label{40}
\begin{eqnarray}
& &J_{xx}=-\frac{2\lambda^{2}(3B+C)}{(U_{0}+\epsilon_{x})^{2}}\Bigl[
-\frac{\sqrt{3}}{U_{0}+\epsilon_{1}}
\frac{t_{0p_{x}}t_{1p_{x}}}{\epsilon_{p_{x}}}
+\frac{1}{\epsilon_{x}}\frac{t_{0p_{x}}^{2}}{\epsilon_{p_{x}}}
\Bigr]^{2}\nonumber\\
&-&\frac{2\lambda^{2}(B+C)}{(U_{0}+\epsilon_{1})^{2}}
\Bigl[\frac{1}{\epsilon_{x}}
\frac{t_{0p_{x}}t_{1p_{x}}}{\epsilon_{p_{x}}}\Bigr]^{2}\nonumber\\
&-&\frac{4\lambda^{2}(-B\sqrt{3})}{(U_{0}+\epsilon_{1})(U_{0}+\epsilon_{x})}
\frac{1}{\epsilon_{x}}\frac{t_{0p_{x}}t_{1p_{x}}}{\epsilon_{p_{x}}}\nonumber\\
& \times &\Bigl[\frac{1}
{\epsilon_{x}}\frac{t_{0p_{x}}^{2}}{\epsilon_{p_{x}}}-
\frac{\sqrt{3}}{U_{0}+\epsilon_{1}}
\frac{t_{0p_{x}}t_{1p_{x}}}{\epsilon_{p_{x}}}\Bigr];\label{40a}
\end{eqnarray}
\begin{eqnarray}
& &J_{yy}=-\frac{2\lambda^{2}(3B+C)}{(U_{0}+\epsilon_{y})^{2}}\Bigl[
\frac{\sqrt{3}}{U_{0}+\epsilon_{1}}\frac{t_{0p_{x}}t_{1p_{x}}}{\epsilon_{p_{x}}}\nonumber\\
&+&\frac{1}{\epsilon_{y}}(\frac{t_{0p_{x}}^{2}}{\epsilon_{p_{x}}}-
\frac{t_{yp_{z}}^{2}}{\epsilon_{p_{z}}})\Bigr]^{2}
-\frac{2\lambda^{2}(B+C)}{(U_{0}+\epsilon_{1})^{2}}\Bigl[\frac{1}{\epsilon_{y}}
\frac{t_{0p_{x}}t_{1p_{x}}}{\epsilon_{p_{x}}}\Bigr]^{2}\nonumber\\
&-&\frac{4\lambda^{2}(B\sqrt{3})}{(U_{0}+\epsilon_{1})(U_{0}+\epsilon_{y})}
\frac{1}{\epsilon_{y}}\frac{t_{0p_{x}}t_{1p_{x}}}{\epsilon_{p_{x}}}\nonumber\\
& \times &\Bigl[\frac{1}
{\epsilon_{y}}(\frac{t_{0p_{x}}^{2}}{\epsilon_{p_{x}}}-\frac{t_{yp_{z}}^{2}}{\epsilon_{p_{z}}})+
\frac{\sqrt{3}}{U_{0}+\epsilon_{1}}\frac{t_{0p_{x}}t_{1p_{x}}}{\epsilon_{p_{x}}}\Bigr];\label{40b}
\end{eqnarray}
\begin{eqnarray}
& &J_{zz}=-\frac{8\lambda^{2}C}{(U_{0}+\epsilon_{z})^{2}}
\Bigl[\frac{1}{\epsilon_{z}}
(\frac{t_{0p_{x}}^{2}}{\epsilon_{p_{x}}}-\frac{t_{zp_{y}}^{2}}
{\epsilon_{p_{y}}})\Bigr]^{2}\nonumber\\
&-&\frac{8\lambda^{2}(4B+C)}{(U_{0}+\epsilon_{1})^{2}}\Bigl[\frac{1}{\epsilon_{z}}
\frac{t_{0p_{x}}t_{1p_{x}}}{\epsilon_{p_{x}}}\Bigr]^{2}.\label{40c}
\end{eqnarray}

\end{mathletters}
\noindent
An alternative derivation of these results is given in the Appendix.

One should keep in mind that Eq. (\ref{25}) holds for a
single bond. To obtain the effective magnetic Hamiltonian of
the entire CuO$_{2}$ plane, one has to sum the magnetic
interaction ${\cal H}(i,j)$ over all bonds, allowing for the
crystal symmetry.  Within a classical approximation the
resulting exchange Hamiltonian of the crystal has only an
easy--plane anisotropy.  To obtain the four--fold anisotropy
within the easy plane requires a consideration of the spin--wave
zero point motion.\cite{PRB}

Now we evaluate these results numerically.  In this connection it
is useful to emphasize that for their less general model
YHAE have shown that the perturbative
results for the anisotropy in the $J$'s agrees to within about
10\% with the numerical evaluations of exact diagonalization
within a Cu--O--Cu cluster.  For our numerical evaluation we
use the parameters of Table 1.  We note that all the nonzero
hopping matrix elements can be expressed in terms of $(pd\sigma )$
and $(pd \pi ) \approx - {1 \over 2} (pd \sigma )$ \cite{LFM}:
\begin{eqnarray}
t_{0,p_x} = - \sqrt 3 t_{1,p_x} = {\sqrt 3 \over 2} (pd \sigma )
\nonumber \\
t_{y,p_z}= t_{0,p_y} = (pd \pi ) \ .
\end{eqnarray}
Thereby we find (in $\mu$eV)
\begin{equation}
\Delta J \equiv J_{\rm av} - J_{\rm zz} = 30 \ ,
\ \ \  \delta J \equiv J_\perp - J_\parallel = 41 \ .
\end{equation}
where $J_{\rm av} = (J_{xx}+ J_{yy})/2$, $J_\perp=J_{yy}$,
and $J_\parallel=J_{xx}$.  Note that the gap in the spin--wave
spectrum due to the easy--plane anisotropy is proportional to
$(\Delta J)^{1/2}$ whereas, as explained by YHAE, the gap
due to the anisotropy within the basal plane is proportional to
$\delta J$. (This statement applies to systems like YB$_2$Cu$_3$O$_6$.
In Sr$_2$CuO$_2$Cl$_2$ the in--plane anisotropy will have contributions
from dipolar interactions.)  In YHAE, the terms in
$J_{xx}$ and $J_{yy}$ involving $B \sqrt 3$ were not included
because only Coulomb matrix elements involving at most two
orbitals were kept.  To see the effect of the additional
terms in the present work, we give, for comparison, the
perturbative results of YHAE: $\Delta J = 30 \mu$eV and
$\delta J = 26 \mu$eV. (The results from exact diagonalization
on a Cu--O--Cu cluster were $\Delta J = 31 \mu$eV and
$\delta J = 31 \mu$eV.)

\section{CONCLUSIONS}
In view of the results of Ref. \onlinecite{ANIS4}, YHAE already demonstrated
that the out--of--plane anisotropy in the superexchange interaction
between Cu ions is dominated by Coulomb exchange terms.  Therefore,
the simplified Coulomb interaction of Eq. (\ref{H0EQ}), used widely
in the literature, is insufficient to explain this anisotropy.
YHAE then considered the anisotropy due to the simplest exchange
terms, like $U_{\alpha \beta \beta \alpha}$, which involve only
two orbitals, and thereby obtained an anisotropy in the
superexchange interaction whose value agreed with experiments.

In the present paper we included all the Coulomb terms which are
allowed by tetragonal site symmetry, and found that the additional
terms, involving $(\alpha 0|\, \alpha 1)$, practically do not affect
the out--of--plane anisotropy $\Delta J$.  The in--plane
single--bond anisotropy, $\delta J$, is somewhat larger than
before.  YHAE showed that the in--plane gap in the spin--wave
spectrum is a manifestation of quantum zero-point fluctuations
and is proportional to $\delta J$.  Using the results of our
present calculation in Eq. (82) of YHAE we estimate the in--plane
gap in the spin--wave spectrum to be about 33
$\mu $eV $\sim$ 0.27cm$^{-1}$.  The direct observation of this
gap would provide an interesting and significant test of our
calculations.

{\bf Acknowledgements}
We acknowledge support from the U. S.--Israel Binational Science
Foundation.  ABH was also supported in part by the U. S. Israel
Education Foundation and by the National Science Foundation under
Grant No. DMR--91--22784.

\begin{appendix}
\section{Alternative calculation of the exchange anisotropy}

In this appendix we obtain the result for $J_{\mu \mu}$ for tetragonal
symmetry by direct application of perturbation theory.  We take
the unperturbed Hamiltonian to be the same as that of Eq. (\ref{H0EQ}),
except that we do not include in it the spin--orbit interaction on
the copper ions. [See Eq. (\ref{CUHAM}).]
That term is now to be included into ${\cal H}_1$.
For simplicity we only consider the contribution to the
anisotropy from channel $a$.  A similar calculation to that given
here shows that channel $b$ gives no anisotropy.  Since we only
consider channel $a$, we can eliminate the oxygens completely
from the problem by introducing an effective hopping matrix
element $\bar t_{\alpha \beta}$ between crystal field states
on near--neighboring copper ions.  For instance,
\begin{eqnarray}
\label{TEQ}
\bar t_{00} & =  & t_{0,p_x}^2/\epsilon_{p_x} \ , \ \
\bar t_{01} = t_{0,p_x}t_{1,p_x}/\epsilon_{p_x} \ , \nonumber
\\ \bar t_{xx} & =  & 0 \ , \ \ \bar t_{yy} = t_{y,p_z}^2/\epsilon_{p_z}
\ , \ \ \bar t_{zz} =  t_{z,p_y}^2/\epsilon_{p_y} \ ,
\end{eqnarray}
as in Eq. (53) of YHAE.  Accordingly, we need to work to fifth
order in perturbation theory, where the perturbations are $V$,
the spin--orbit interaction, which we take to second order,
${\cal H}_{\rm hop} \equiv {\cal T}$, the hopping between copper
ions, which we take to second order, and
$\Delta {\cal H}_C={\cal C}$ on the copper ions, which we
take to first order.  We also use the fact that
$U_{\alpha \beta \gamma \delta}$ is only nonzero when
$\sigma(\alpha) \sigma(\beta) \sim \sigma(\gamma) \sigma(\delta)$.

In fifth order perturbation theory there are, a priori, 30
different ways to order these perturbations.  But obviously
the Coulomb perturbation can only appear when the two holes
are on the same site.  So we have only to consider how to
insert two powers of $V$ into the sequence ${\cal T}{\cal C} {\cal T}$.
There are basically three types of terms to consider.  The first is
\begin{equation}
\label{PERT}
H_1 \equiv \left( V {1 \over {\cal E} } {\cal T}
+ {\cal T} {1 \over {\cal E} } V \right)
{1 \over {\cal E} } {\cal C} {1 \over {\cal E} }
\left( V {1 \over {\cal E} } {\cal T} + {\cal T} {1 \over {\cal E} } V
\right) \ ,
\end{equation}
where $\cal E$ denotes the appropriate energy denominator.
The second type of terms are those in which the two $V$'s
are both to the left of ${\cal C}$, and the third type are terms
which are the Hermitian conjugates of the second type, i. e. those
in which the two $V$'s are to the right of ${\cal C}$.

That the second type of term does not lead to any anisotropy
can be established by the following argument.  Suppose we show
that these terms vanish when applied to any triplet spin state.
That would imply that these terms are of the form
$(1/4) - {\bf S}_i \cdot {\bf S}_j$, which obviously gives rise to
no anisotropy.  The second type of term has first (reading from
right to left) ${\cal T}$ acting on the triplet state.  That will put
the two holes, which were initially distributed one
on site $i$, the other on site $j$, onto the same site, perforce
one in state $|\, 0\rangle$, the other in state $|1\rangle$.  Now
apply the Coulomb perturbation.  This operator can leave the
two holes in the same states, viz. one in $| \, 0\rangle$ and the
other in $|1\rangle$.  In fact, by our observation on the
form of $U_{\alpha \beta \gamma \delta}$ and by the fact that
parallel spins are not allowed in the same spatial orbital, there
are no other final states. But such a diagonal matrix element
of $\bf U$ was treated by YHAE and found to give no anisotropy
in the second type of term.  So we conclude that the second
type of term does not give rise to any anisotropy.  Terms
of the third type are the Hermitian conjugate of type two and
therefore are subject to the same argument.  Thus all the
anisotropic terms are contained in the expression in Eq. (\ref{PERT}).

If $T_{ij}$ denotes hopping from $i$ to $j$, we can write
Eq. (\ref{PERT}) as
\begin{equation}
H_1 = 2 \sum_{\alpha=x,y,z} {\bf Q}_\alpha^\dagger {1 \over {\cal E}} C
{1 \over {\cal E}} {\bf Q}_\alpha \ ,
\end{equation}
where
\begin{equation}
{\bf Q}_\alpha = \left( V_\alpha {1 \over {\cal E}} T_{ij}
+ T_{ij}  {1 \over {\cal E}} V_\alpha \right) \ .
\end{equation}
Here $V_\alpha = \sum_h L_\alpha (h) s_\alpha (h)$, where the
sum is over the two holes, $h$.  Acting on the ground state,
the operator ${\bf Q}_\alpha$ produces two final states, depending
on whether the orbital $|0 \rangle$ or $|1\rangle$ is occupied.  So
we define
\begin{eqnarray}
&& [Q_\alpha^{(\gamma)} ]_{\sigma , \eta ; \sigma' ; \eta'} =
\langle 0| d_{i, \gamma, \sigma} d_{i, \alpha ,\eta}
\nonumber \\
&& \ \ \ \times \left( V_\alpha {1 \over {\cal E}} T_{ij}
+ T_{ij}  {1 \over {\cal E}} V_\alpha \right) d^\dagger_{i,0,\eta'}
d^\dagger_{j,0,\sigma'} | 0 \rangle \ ,
\end{eqnarray}
where $\gamma$ can be 0 or 1.  Then
\begin{equation}
H_1 = 2 \sum_{\alpha \gamma \gamma'} \left[ {\bf Q}_\alpha^{(\gamma')}
\right]^\dagger \left[ {\cal C}_\alpha^{\gamma' \gamma}
\right] \left[ {\bf Q}_\alpha^{(\gamma)} \right] \Delta_{\gamma\alpha}^{-1}
\Delta_{\gamma'\alpha}^{-1} \ ,
\end{equation}
where $\Delta_{\gamma\alpha} = \epsilon_\gamma+ \epsilon_\alpha + U_0$
and
\begin{eqnarray}
\label{APP1}
 \left[ {\cal C}_\alpha^{\gamma' \gamma} \right]_{\sigma'' \eta'' ;
\sigma' \eta' } & = & \langle 0| d_{i \gamma' \sigma''} d_{i \alpha \eta''}
\Delta {\cal H}_C
d^\dagger_{i \alpha \eta'} d^\dagger_{i \gamma \sigma'}
|0 \rangle \nonumber \\ &=&
\delta_{\eta'', \eta'} \delta_{\sigma' \sigma''} \langle \alpha \gamma'
| \Delta {\cal H}_C | \alpha \gamma  \rangle \nonumber \\
&& -  \delta_{\eta'', \sigma'}
\delta_{\sigma'' , \eta'} \langle \alpha \gamma' | \Delta {\cal H}_C
| \gamma \alpha \rangle \ .
\end{eqnarray}
Now we introduce the notation for direct products
\begin{equation}
[ {\cal A} {\cal B}]_{\sigma' , \eta' ; \sigma , \eta}
= A_{\sigma' \sigma} B_{\eta' \eta} \ ,
\end{equation}
so that
\begin{equation}
{\cal C}_\alpha^{\gamma' , \gamma} = [ {\cal O } ]
\langle \alpha \gamma' | \Delta {\cal H}_C | \alpha \gamma  \rangle
\equiv [ {\cal O } ] ( \alpha \gamma | \, \alpha \gamma' )
\end{equation}
in the notation of Eqs. (\ref{COULOMB}).
Also $\left[ {\cal O} \right] = [ {\cal I} {\cal I}
- \vec \sigma \cdot \vec \sigma ]/2$.

 From Appendix H of YHAE we also take the results
\begin{equation}
\left[ {\bf Q}_\alpha^{(0)} \right] =
(C_1 + C_2 ) \left[ {\cal I} \sigma_\alpha \right]
- C_2 \left[ {\cal I} \sigma_\alpha \right] \left[ {\cal O} \right]
\end{equation}
and
\begin{equation}
\left[ {\bf Q}_\alpha^{(1)} \right] = C_3
\left[ {\cal I } \sigma_\alpha \right] \ ,
\end{equation}
where
%\begin{equation}
%C_1 = {\lambda \over 2 } \Biggl\{
%{\bar t_{00} \langle \alpha | L_x | 0 \rangle \over \epsilon_\alpha }
%+ {\bar t_{00} \langle \alpha | L_x | 0 \rangle \over U_0 } \Biggr\} \ ,
%\end{equation}
%\begin{equation}
%C_2 = - { \lambda \over 2 } \Biggl\{
%{\bar t_{00} \langle \alpha | L_x | 0 \rangle \over U_0 }
%+  {\bar t_{\alpha \alpha} \langle \alpha | L_\alpha | 0 \rangle
%\over \epsilon_\alpha }
%+  {\bar t_{01} \langle \alpha | L_\alpha | 1 \rangle
%\over (\epsilon_1 + U_0 ) } \Biggr\}  \ .
%\end{equation}
\begin{equation}
\label{C12}
C_1+C_2 = \left[ {(\bar t_{\alpha \alpha } - \bar t_{00}){\cal L}_{0\alpha}
\over \epsilon_\alpha } + {\bar t_{01} {\cal L}_{1 \alpha} \over
( \epsilon_1+U_0 ) } \right]
\end{equation}
and
\begin{equation}
\label{C3}
C_3 = - \bar t_{01} {\cal L}_{0 \alpha }
\left( {\epsilon_1 + \epsilon_\alpha + U_0
\over \epsilon_\alpha (\epsilon_1 + U_0) } \right) \ .
\end{equation}

Thus we obtain the result
\begin{eqnarray}
{\cal H}(i,j) && = \sum_\alpha \Biggl\{
{ 2 (\alpha 0 | \, \alpha 0 ) \over
\Delta_0^2 } \left[ C_1^* {\cal I} {\cal I} + C_2^* {\cal I} {\cal I}
- C_2^* {\cal O} \right] \nonumber \\ && \times \ \
\left[ {\cal I} \sigma_\alpha \right]
\left[ {\cal O} \right] \left[  {\cal I} \sigma_\alpha
\right] \left[ C_1 {\cal I} {\cal I} + C_2 {\cal I} {\cal I}
- C_2 {\cal O} \right] \nonumber \\
&& + { 2(\alpha 1 | \, \alpha 1 ) \over \Delta_1^2 }
( C_3^* \left[ {\cal I} \sigma_\alpha \right] \left[
{\cal O} \right] \left[ {\cal I} \sigma_\alpha
\right] C_3 \Bigr) \nonumber \\
&& + { 2(\alpha 0 | \, \alpha 1 ) \over \Delta_0 \Delta_1 }
C_3^* \left[ {\cal I} \sigma_\alpha \right] \left[ {\cal O} \right]
\nonumber \\ && \times \left[ {\cal I} \sigma_\alpha \right]
\left[ C_1 {\cal I} {\cal I} + C_2 {\cal I} {\cal I} - C_2 {\cal O} \right]
\nonumber \\
&& + { 2(\alpha 0 | \, \alpha 1 ) \over \Delta_0  \Delta_1 }
\left[ C_1^* {\cal I} {\cal I} + C_2^* {\cal I} {\cal I}
- C_2^* {\cal O} \right] \nonumber \\ && \times
\left[ {\cal I} \sigma_\alpha \right] \left[ {\cal O} \right]
\left[ {\cal I} \sigma_\alpha \right] C_3 \Biggr\} \ .
\end{eqnarray}
Now we use the equality
\begin{equation}
\left[ {\cal I} \sigma_\alpha \right] \left[ {\cal O}
\right] \left[ {\cal I} \sigma_\alpha \right]
= \frac{1}{2} \left[ {\cal I} {\cal I} - 2 \sigma_\alpha \sigma_\alpha
+ \vec \sigma \cdot \vec \sigma  \right]  \ .
\end{equation}
So, dropping isotropic terms, we have \cite{NOTE}
\begin{eqnarray}
&& {\cal H}(i,j) = \sum_\alpha \Biggl\{
- { 2 (\alpha 0 | \, \alpha 0 ) \over
\Delta_{0\alpha}^2 } \left[ C_1^* {\cal I} {\cal I}
+ C_2^* {\cal I} {\cal I} - C_2^* {\cal O} \right] \nonumber \\
&\times& \ \  \left[
\sigma_\alpha \sigma_\alpha \right] \left[ C_1 {\cal I} {\cal I}
+ C_2 {\cal I} {\cal I} - C_2 {\cal O} \right] \nonumber \\
&-& { 2 (\alpha 1 | \, \alpha 1) \over \Delta_{1\alpha}^2 }
C_3^* \left[ \sigma_\alpha \sigma_\alpha \right] C_3 \nonumber \\
& -& { 2 (\alpha 0 | \, \alpha 1) \over \Delta_{0\alpha}
\Delta_{1\alpha}} C_3^* \left[ \sigma_\alpha \sigma_\alpha \right]
\left[ C_1 {\cal I} {\cal I} + C_2 {\cal I} {\cal I}
- C_2 {\cal O} \right] \nonumber \\
& - &{ 2 (\alpha 1 | \, \alpha 0 ) \over \Delta_{0\alpha}
\Delta_{1\alpha} } \left[ C_1^* {\cal I} {\cal I} + C_2^* {\cal I} {\cal I}
- C_2^* {\cal O} \right] \left[ \sigma_\alpha \sigma_\alpha \right]
C_3 \Biggr\} \ .
\end{eqnarray}

Now use
\begin{equation}
2 \left[ \sigma_\alpha \sigma_\alpha \right] \left[ {\cal O} \right]
= 2 \left[ {\cal O} \right] \left[ \sigma_\alpha \sigma_\alpha \right]
= \left[ \vec \sigma \cdot \vec \sigma - {\cal I} {\cal I} \right] \ .
\end{equation}
Thus the terms involving ${\cal O}$ are isotropic.  So the
anisotropic terms are correctly given by Eq. (\ref{27}) with
\begin{eqnarray}
J_{\mu \mu } = & - & { 8 (\mu 0 | \, \mu 0 ) \over
\Delta_{0\mu}^2 } |C_1 + C_2|^2 - {8 (\mu 1 | \, \mu 1 )
\over \Delta_{1\mu }^2 } |C_3|^2 \nonumber \\
& - & 16 {\rm Re} { (\mu 0 | \, \mu 1) \over \Delta_{0\mu}
\Delta_{1\mu}} C_3^* (C_1+C_2) \ .
\end{eqnarray}
Now we use Eqs. (\ref{C12}), and (\ref{C3}) to write
\begin{eqnarray}
J_{\mu \mu } =
&& { 8 (\mu 0| \, \mu 0 ) \over (U_0 + \epsilon_\mu)^2}
\left[ { (\bar t_{00}-\bar t_{\mu \mu} ) {\cal L}_{0 \mu } \over
\epsilon_\mu} - { \bar t_{01} {\cal L}_{1 \mu } \over (\epsilon_1+U_0)
} \right]^2 \nonumber \\
&& + { 8 (\mu 1 | \, \mu 1) \over \epsilon_\mu^2
( \epsilon_1+U_0)^2 } \bar t_{01}^2 {\cal L}_{0\mu}^2
- {16 (\mu 0| \, \mu 1) \bar t_{10} {\cal L}_{0 \mu}
\over  \epsilon_\mu (U_0+\epsilon_\mu )(\epsilon_1 +U_0)} \nonumber \\
&& \times
\left[ {(\bar t_{\mu \mu } - \bar t_{00}){\cal L}_{0\mu}
\over \epsilon_\mu } + {\bar t_{01} {\cal L}_{1 \mu } \over
( \epsilon_1+U_0 ) } \right] \ .
\end{eqnarray}
This result reproduces Eqs. (\ref{40}) of the text.
\end{appendix}

%   lines and directly read in your .bbl file if you use bibtex.

% \begin{figure}
% \caption{}
% \label{}
% \end{figure}

\begin{table}
\caption{Values (in eV) of the parameters used to calculate the anisotropic
exchange.  For a discussion of these values, see YHAE.}
\label{TAB1}
\begin{tabular}{c c c c c c c c c c c}
$\lambda$ & $A$ & $B$ & $C$ & $(pd \sigma)$ & $\epsilon_1$ &
$\epsilon_x = \epsilon_y$ & $\epsilon_z$ & $\epsilon_{p_x}$ &
$\epsilon_{p_y}$ & $\epsilon_{p_z}$ \\
0.1 & 7.0 & 0.15 & 0.58 &  1.5 & 1.8 & 1.8 & 1.8 & 3.25 & 3.25 & 3.25 \\
\end{tabular}
\end{table}

\begin{table}
\caption{Values of
$(\alpha \delta | \beta \gamma )= U_{\alpha \beta \gamma \delta}$
in terms of the Racah coefficients, taken\protect \cite{NOTE}
from Table A26 of Ref. \protect\onlinecite{GRIFF}).}
\label{TAB2}
\begin{tabular}{c c c c }
& $\alpha=x$ & $\alpha=y$ & $\alpha=z$ \\
($\alpha 0 | \alpha$ 0) = & 3B+C & 3B+C & C  \\
($\alpha 0 | \alpha$ 1) = & $- \sqrt 3$ B & $\sqrt 3$ B & 0 \\
($\alpha 1 | \alpha$ 1) = & B+C & B+C & 4B+C  \\
\end{tabular}
\end{table}


\begin{references}
\bibitem{NOTE}
The states $\epsilon$, $\theta$, $\xi$, $\eta$, and $\zeta$ in Table A26
of Ref. \onlinecite{GRIFF} should be identified with our states $| \,0\rangle$,
$|1\rangle$, $|x\rangle$, $|y\rangle$, and $|z\rangle$, respectively.
\end{references}
\end{document}